\documentclass[final,5p,times,twocolumn]{elsarticle}
\makeatletter
\def\ps@pprintTitle{%
 \let\@oddhead\@empty
 \let\@evenhead\@empty
 \def\@oddfoot{}%
 \let\@evenfoot\@oddfoot}
\makeatother

\usepackage{graphicx}
\usepackage{amssymb}
\usepackage{mathtools, cuted}

\usepackage{lineno}

\usepackage{amsmath}
\usepackage{bm}
\usepackage{color}

\journal{Computational Materials Science}

\begin{document}

\begin{frontmatter}


\title{First-principles calculations for attosecond electron dynamics in solids}



 \author[tsukuba,mpsd]{Shunsuke A. Sato}
 \ead{ssato@ccs.tsukuba.ac.jp}
 \address[tsukuba]{Center for Computational Sciences, University of Tsukuba, Tsukuba 305-8577, Japan}
 \address[mpsd]{Max Planck Institute for the Structure and Dynamics of Matter, Luruper Chaussee 149, 22761 Hamburg, Germany}


\begin{abstract}
Nonequilibrium electron dynamics in solids is an important subject from both fundamental and technological points of view. The recent development of laser technology has enabled us to study ultrafast electron dynamics in the time domain. First-principles calculation is a powerful tool for analyzing such complex electron dynamics and clarifying the physics behind the experimental observation. In this article, we review the recent development of the first-principles calculation for light-induced electron dynamics in solids by revising its application to recent attosecond experiments. The electron dynamics calculations offer an accurate description of static and transient optical properties of solids and provide physics insight into light-induced electron dynamics. Furthermore, the microscopic decomposition of transient properties of nonequilibrium systems has been developed to extract microscopic information from the simulation results. The first-principles analysis opened a novel path to analyze the nonequilibrium electron dynamics in matter and to provide the fundamental understanding complementarily with the sophisticated experimental technique.
\end{abstract}


\end{frontmatter}



\section{Introduction \label{sec:intro}}

Interaction of light with matter is one of the most important subjects in physics, and it has been intensively studied \cite{MOUROU2012720,Krausz2014,Basov2017}. Recent developments of laser technologies enabled us to study interactions of strong light with matter, and nonequilibrium electron dynamics in matter have been investigated \cite{RevModPhys.72.545,RevModPhys.81.163}. Once very strong light is irradiated on solids, highly-nonlinear electron dynamics are induced, triggering various intriguing phenomena such as high-order harmonic generation \cite{Ghimire2011,Ghimire2019}, nonlinear current induction \cite{Schiffrin2013,Higuchi2017}, and laser ablation \cite{Gattass2008,Balling_2013}. These highly-nonlinear phenomena may further open technological applications such as novel light source, ultrafast optoelectronic devices, and fine laser-machining. However, to realize such applications, the fundamental understanding of the nonequilibrium electron dynamics in matter is indispensable.

After observing the high-order harmonic generation from noble gases \cite{McPherson:87,Ferray_1988}, experimental techniques to generate very short laser pulses with attosecond time duration have been rapidly developed \cite{RevModPhys.81.163}. The generated attosecond laser pulses have been employed to investigate electron dynamics in atoms and molecules in the time domain \cite{Goulielmakis2010,PhysRevLett.105.143002,PhysRevLett.106.123601}. Recently, the attosecond experimental technique has been further applied to solid-state materials \cite{Schultze1348,Lucchini916,Moulet1134,Siegrist2019,buades2020attosecond,lucchini2020unravelling}, enabling us to explore light-induced nonequilibrium electron dynamics in solids. However, reflecting the complex electronic structure of solids, the attosecond spectroscopy for solids tends to provide complex experimental data. Hence it is not straightforward to interpret or understand the experimental observation. Therefore, to access the physics behind the obtained experimental result, a complementary theoretical investigation is often a key.

First-principles calculations based on the time-dependent density functional theory (TDDFT) \cite{PhysRevLett.52.997} is a powerful tool to investigate such complex electron dynamics in solids \cite{PhysRevB.62.7998,PhysRevB.82.155110,PhysRevB.85.045134} because it describes the electron dynamics based on the \textit{ab-initio} electron-ion interacting Hamiltonian without empirical parameters for the description of electron dynamics. The simulations based on TDDFT has been applied to nonlinear light-matter interactions and provides microscopic insight into the highly-nonlinear light-induced phenomena such as the high-order harmonic generation \cite{PhysRevLett.118.087403,PhysRevA.97.011401}, laser-ablation \cite{PhysRevB.92.205413,10.1117/12.2243012}, nonlinear current injection \cite{PhysRevLett.113.087401,Wachter_2015}, and formation of Floquet-Bloch states \cite{DeGiovannini2016,Huebener2017}.

By extending the TDDFT electron dynamics simulation, we developed the theoretical and numerical schemes to study the attosecond electron dynamics in solids. The combination of the \textit{ab-initio} electron dynamics simulation and the attosecond transient spectroscopy has opened a novel avenue to explore nonequilibrium electron dynamics in matter and has provided the fundamental insight into the nonequilibrium electron dynamics \cite{Schultze1348,Lucchini916,Sommer2016,Schlaepfer2018,Volkov2019,Lucchini_2020}. In this article, we review the recent development of the first-principles calculation for attosecond electron dynamics in solids and the theoretical schemes to extract microscopic information from the simulations.

The paper is organized as follows: in section~\ref{sec:theory}, we first review the theoretical framework to describe electron dynamics in solids on the basis of the time-dependent density functional theory. In section~\ref{sec:numerics}, the numerical implementation to practically perform the first-principles electron dynamics simulation is reviewed. In section~\ref{sec:rt-dynamcis}, we review the first-principles electron dynamics calculation and its application to the recent attosecond experiments. Finally, the summary will be presented in Sec.~\ref{sec:summary}.
Hereafter, the atomic units are used unless stated otherwise.

\section{Theoretical framework \label{sec:theory}}

Linear and low-order nonlinear optical properties of solids have been theoretically investigated with the perturbation theory in the frequency domain \cite{PhysRevB.55.4343,PhysRevLett.80.3320,PhysRevB.62.4927,PhysRevLett.88.066404,PhysRevLett.107.186401,PhysRevB.53.15638}. The perturbative analysis provides a deep insight into the optical phenomena. However, the perturbative approaches face conceptual and practical difficulties in describing highly-nonlinear phenomena and recent time-domain experiments. A time-domain simulation would be a straightforward approach to investigate such highly-nonlinear phenomena and real-time dynamics because it does not rely on the perturbative expansion but naturally capture the time propagation of quantum systems in the time domain. Here, we describe the theoretical framework of the real-time electron dynamics simulations in solids on the basis of the time-dependent density functional theory (TDDFT) \cite{PhysRevLett.52.997}.

The TDDFT offers a formally exact description of time-dependent electron density $\rho(\bm r,t)$ through the time-evolution of an auxiliary system, the so-called time-dependent Kohn-Sham system. To introduce TDDFT, let us start from the time-dependent Schr\"odinger equation for an $N$-electron system interacting with ions under the presence of one-body external potential $v(\bm r, t)$
\begin{align}
i \frac{\partial}{\partial t}\Psi(\bm r_1, \cdots, \bm r_N, t) = \hat H(t) \Psi(\bm r_1, \cdots, \bm r_N, t),
\label{eq:tdse}
\end{align}
with the many-body Hamiltonian $\hat H(t)$ given by
\begin{align}
\hat H(t) = \sum^N_j \left [ \frac{\bm p^2_j}{2} + v_{ion}(\bm r_j) + v_{ext}(\bm r_j, t) \right ]
+ \sum^N_{j>k} \frac{1}{\left |\bm r_j - \bm r_k \right |},
\end{align}
where $v_{ion}(\bm r)$ is the electron-ion interacting potential. It is given by
\begin{align}
v_{ion}(\bm r) = - \sum_{a} \frac{Z_a}{\left | \bm r - \bm R_a \right |}.
\end{align}
Here, $Z_a$ is the ionic charge and $\bm R_a$ is the ionic position. In this article, we assume that the ionic positions are frozen since we are interested in only ultrafast electron dynamics within a very short timescale.

By solving the time-dependent Schr\"odinger equation, Eq.~(\ref{eq:tdse}), one can evaluate the time-dependent electron density as
\begin{align}
\rho(\bm r,t)= N \int d\bm r_2 \cdots d \bm r_N \left |\Psi(\bm r, \bm r_2, \cdots, \bm r_N, t) \right |^2.
\end{align}
Therefore, the time-dependent Schr\"odinger equation can be seen as a mapping from the external potential $v_{ext}(\bm r, t)$ to the density $\rho(\bm r,t)$: $v_{ext}(\bm r,t)\rightarrow \rho(\bm r, t)$. Remarkably, the Runge-Gross theorem \cite{PhysRevLett.52.997}, which is the central theorem of TDDFT, shows that the mapping is invertible, and there is a one-to-one correspondence between the external potential $v_{ext}(\bm r, t)$ and the density $\rho(\bm r,t)$. Based on the one-to-one correspondence of the external potential and the density, one can introduce an auxiliary \textit{non-interacting} system that provides the identical density to the original many-body system. The auxiliary system is the so-called time-dependent Kohn-Sham system. Since the time-dependent Kohn-Sham system is a non-interacting system, it can be described by a single Slater determinant. Each orbital of the Slater determinant obeys a one-body Schr\"odinger-like equation, which is the so-called time-dependent Kohn-Sham equation,
\begin{align}
i \frac{\partial}{\partial t}\psi_j(\bm r, t) =
\left[\frac{\bm p^2}{2} + v_{KS}(\bm r, t) \right ]\psi_j(\bm r, t),
\label{eq:tdks-00}
\end{align}
where $\psi_j(\bm, t)$ is a Kohn-Sham orbital, $j$ is an orbital index, and $v_{KS}(\bm r, t)$ is a one-body potential introduced such that the Kohn-Sham system reproduces the identical density to the original many-body system as 
\begin{align}
\rho(\bm r, t) = \sum^N_j \left |\psi_j(\bm r, t) \right |^2 = N \int d\bm r_2 \cdots \bm r_N \left |\Psi(\bm r, \bm r_2, \cdots, \bm r_N, t) \right |^2.
\label{eq:density}
\end{align}

The one-body potential $v_{KS}(\bm r,t)$, which is the so-called Kohn-Sham potential, consists of several components as
\begin{align}
v_{KS}(\bm r, t) = v_{ion}(\bm r) + v_{H}(\bm r,t)  + v_{XC}(\bm r,t) + v_{ext}(\bm r,t),
\label{eq:ks-pot}
\end{align}
where $v_{H}(\bm r, t)$ is the Hartree potential given by
\begin{align}
v_H(\bm r, t) = \int d\bm r' \frac{\rho(\bm r', t)}{\left |\bm r - \bm r' \right |},
\end{align}
and $v_{xc}(\bm r, t)$ is the exchange-correlation potential that describes all the rest of the Kohn-Sham potential. Thus, once the exact exchange-correlation potential is given, the time-dependent Kohn-Sham equation, Eq.~(\ref{eq:tdks-00}), provides the exact electron dynamics. However, the exact exchange-correlation potential is an unknown object in TDDFT, as well as the static density functional theory (DFT). Therefore, the exchange-correlation potential has to be approximated in practical calculations. Properties of exchange-correlation potentials (functionals) have been intensively studied \cite{PhysRevA.49.2421,PhysRevA.30.2745,PhysRevA.29.2322,PhysRevA.40.4190,doi:10.1063/1.3271392,PhysRevB.93.155146}, and various approximated potentials (functionals) have been developed \cite{PhysRevB.23.5048,PhysRevB.45.13244,PhysRevA.49.2421,PhysRevLett.77.3865,PhysRevLett.91.146401,doi:10.1063/1.476577,PhysRevLett.115.036402,doi:10.1063/1.472933,doi:10.1063/1.1564060}. Thanks to the development of sophisticated potentials (functionals), the quality of the TDDFT simulation has been significantly improved \cite{PhysRevB.78.121201,PhysRevLett.107.216402,doi:10.1063/1.4937379}. Furthermore, the development of improved approximations is still one of the central fields of TDDFT and DFT, and fundamental studies still continue presently \cite{PhysRevLett.109.266404,doi:10.1063/1.4867002,PhysRevLett.109.036402,PhysRevLett.119.263401}.

Here, we review a specific form of the time-dependent Kohn-Sham equation to simulate light-induced electron dynamics in solids practically \cite{PhysRevB.62.7998,PhysRevB.89.064304}. For this purpose, we first assume that the wavelength of laser fields is much longer than the spatial scale of electron dynamics. Hence the electric fields can be approximated by spatially homogeneous fields. This is the so-called \textit{dipole approximation}, and the dipole-field simply describes the laser field contribution in the time-dependent Kohn-Sham equation, Eq.~(\ref{eq:tdks-00}), as
\begin{align}
v_{ext}(\bm r, t) = \bm E(t)\cdot \bm r,
\label{eq:pot-length-gauge}
\end{align}
where $\bm E(t)$ is the applied electric field. This is one of the possible expressions of the light-matter interactions in the Schr\"odinger equation, and it is called \textit{length gauge}. The length gauge has been widely employed to investigate light-induced electron dynamics in isolated systems such as atoms, molecules, and clusters in both linear and nonlinear regimes \cite{PhysRevB.54.4484,Kawashita_2009,Sato2018}. However, the external potential in the length gauge, Eq.~(\ref{eq:pot-length-gauge}), breaks the periodicity of the Hamiltonian of crystalline solids. Thus, the direct use of the length gauge in numerical simulations is not feasible. To obtain a Hamiltonian that has compatible symmetries with solids, one can apply the following gauge transformation $\psi_j (\bm r, t) = e^{i\bm A\cdot \bm r} \tilde \psi_j (\bm r, t)$ and rewrite the time-dependent Kohn-Sham equation as
\begin{align}
i\frac{\partial}{\partial t} \tilde \psi_j (\bm r, t) =& \Bigg [ 
\frac{1}{2}\left (\bm p + \bm A \left (t \right ) \right )^2 + e^{-i\bm A(t)\cdot \bm r} v_{ion}(\bm r, t)e^{i\bm A(t)\cdot \bm r}
 \nonumber \\
&+ v_H(\bm r, t) + v_{xc}(\bm r,t)
\Bigg ]\tilde \psi_j(\bm r,t),
\label{eq:tdks-velocity}
\end{align}
where the vector potential $\bm A(t)$ is related to the electric field by
\begin{align}
\bm A(t) = - \int^t dt' \bm E(t').
\end{align}
Note that, if the ionic potential $v_{ion}(\bm r)$ in Eq.~(\ref{eq:tdks-velocity}) is a spatially local operator, it commutes with the phase factor $e^{i\bm A\cdot \bm r}$, and the phase factors are canceled out in the Kohn-Sham equation, Eq.~(\ref{eq:tdks-velocity}). However, in practical calculations, the ionic potentials are treated with the pseudopotential approximation \cite{PhysRevLett.48.1425,PhysRevB.43.1993,OLIVEIRA2008524}. Thus, they can be a spatially nonlocal operator in general. Hence, the phase factors $e^{i\bm A\cdot \bm r}$ need to be treated explicitly in practical calculations.

In Eq.~(\ref{eq:tdks-velocity}), the contribution of the electric field is described by the vector potential $\bm A(t)$. This choice of the gauge degree of freedom is called \textit{velocity gauge}, and it has been applied to investigate light-induced electron dynamics in solids \cite{PhysRevB.62.7998}. In Eq.~(\ref{eq:tdks-velocity}), the Hamiltonian has the same periodicity as the ionic potential $v_{ion}(\bm r)$, having the same symmetry as the crystal. Thus, one may employ the following time-dependent Bloch ansatz
\begin{align}
\tilde \psi_j(\bm r, t) = e^{i\bm k \cdot \bm r}u_{b \bm k}(\bm k, t),
\end{align}
where the Bloch orbitals $u_{b \bm k}(\bm k, t)$ have the same periodicity as the ionic potential $v_{ion}(\bm r)$, and they are labeled with the band index $b$ and the Bloch wavevector $\bm k$ instead of the orbital index $j$. One may further obtain the time-dependent Kohn-Sham equation for the Bloch orbitals as
\begin{align}
i \frac{\partial}{\partial t} u_{b\bm k}(\bm r, t) = \hat h_{KS,\bm k}(t) u_{b\bm k}(\bm r, t),
\label{eq:tdks-velocity-bloch}
\end{align}
where the one-body Hamiltonian is given by
\begin{align}
\hat h_{KS,\bm k}(t) = & 
\frac{1}{2}\left (\bm p + \bm k + \bm A(t) \right )^2 + e^{-i\left (\bm A(t) + \bm k \right )\cdot \bm r} v_{ion}(\bm r, t)e^{i \left ( \bm A(t) + \bm k \right )\cdot \bm r}
 \nonumber \\
&+ v_H(\bm r, t) + v_{xc}(\bm r,t).
\label{eq:tdks-velocity-bloch-ham}
\end{align}
This is our central equation of motion to describe light-induced electron dynamics in solids.

One of the central physical quantities evaluated with the electron dynamics calculation is the electric current since it is the source term of Maxwell's equation and determines the optical properties of matter. To compute the electric current, we first define the kinetic momentum as
\begin{align}
\bm \pi_{\bm k}(t) &= \frac{1}{i} \left [\bm r,  \hat h_{KS,\bm k}(t)\right ]  \nonumber \\
&= \bm p + \bm k  + \bm A(t) + \frac{1}{i}
\left [\bm r,  e^{-i\left (\bm A(t) + \bm k \right )\cdot \bm r} v_{ion}(\bm r, t)e^{i \left ( \bm A(t) + \bm k \right ) \cdot \bm r}  \right ],
\end{align}
where $\left [ A, B \right ]$ is the commutator defined as $\left [ A, B \right ] = AB-BA$.
Then, one can evaluate the electric current density as
\begin{align}
\bm J(t) = -\frac{1}{\Omega} \sum_{b} \int_{BZ} d\bm k f_{b\bm k} \int_{\Omega} d \bm r
u^*_{b\bm k}(\bm r, t) \bm \pi_{\bm k}(t) u_{b\bm k}(\bm r, t),
\label{eq:current-density}
\end{align}
where $\Omega$ is the volume of the unit-cell, and $f_{b\bm k}$ is the occupation factor of each orbital.

In summary, to investigate light-induced electron dynamics in solids, we employ the time-dependent Kohn-Sham equation for Bloch orbitals, Eq.~(\ref{eq:tdks-velocity-bloch}). The primary input of the electron dynamics simulation is the vector potential $\bm A(t)$, or, equivalently, the electric field $\bm E(t)$. The main output of the simulation is the electric current evaluated with Eq.~(\ref{eq:current-density}). Thus, the electron dynamics simulation can be seen as a numerical constitutive relation between the electric field $\bm E(t)$ and the induced current $\bm J(t)$, and it can be used to extract linear and nonlinear optical properties of matter.

\section{Numerical implementation \label{sec:numerics}}

To practically solve the time-dependent Kohn-Sham equation, Eq.~(\ref{eq:tdks-velocity-bloch}), we employ two numerical approaches, depending on computational costs and nature of dynamics. One is the real-space grid representation \cite{PhysRevC.17.1682,PhysRevB.50.11355,PhysRevB.54.4484,NODA2019356,doi:10.1063/1.5142502}, and the other is the orbital basis expansion \cite{PhysRevB.89.224305}. Here, we briefly review the two approaches. Note that, once converged simulation results are obtained, the results are independent from the numerical representation as the accurate solutions of Eq.~(\ref{eq:tdks-velocity-bloch}) are obtained.

\subsection{Real-space representation \label{subsubsec:real-space}}

First, we revisit the real-space grid representation. In this approach, the wavefunctions are represented on discretized grid points as
\begin{align}
u_{b\bm k}(\bm r, t) \rightarrow u_{b\bm k}(\bm r_j, t),
\end{align}
where an integer $j$ represents a serial number of spatial grids in a unit cell. Here, we consider that the unit cell is discretized into $N_g$ grid points.

In the real-space grid representation, the operation of a general operator $\hat O$ to a Bloch orbital $u_{b\bm k}(\bm r,t)$ is described by the following matrix operation
\begin{align}
  \left [\hat O  u_{b\bm k}(\bm r,t) \right ]_{\bm r = \bm r_i} = \sum^{N_g}_{j=1} O_{ij} u_{b \bm k}(\bm r_j, t),
\end{align}
where $O_{ij}$ is a matrix representation of the operator $\hat O$ in the grid representation. Note that spatially-local operators become diagonal matrices in the grid representation. For example, a local potential $v_L(\bm r)$ is represented as $\left [v_L(\bm r) \right ]_{ij} = \delta_{ij} v_L(\bm r_j)$. In the grid representation, the Kohn-Sham Hamiltonian in Eq.~(\ref{eq:tdks-velocity-bloch-ham}) can be expressed as
\begin{align}
  \left [ \hat h_{KS,\bm k}(t)\right ]_{ij} =& -\frac{C_{ij}}{2} +\frac{\bm g_{ij}}{i} \cdot  \left (\bm k + \bm A(t) \right )
+ e^{-i  \bm A(t)\cdot \bm r_i} v_{ion, ij} e^{i \bm A(t)\cdot \bm r_j} \nonumber \\
& +\delta_{ij} \left [ \frac{1}{2} \left (\bm k + \bm A(t) \right )^2  +v_{H}(\bm r_j,t) + v_{xc}(\bm r_j, t) \right ],
\label{eq:ham-matrix-grid}
\end{align}
where $C_{ij}$ and $\bm g_{ij}$ are finite-difference coefficients for the Laplacian $\nabla^2$ and the gradient $\bm \nabla$, respectively. Note that the finite-difference coefficient matrices, $C_{ij}$ and $\bm g_{ij}$, can be chosen as sparse matrices \cite{PhysRevB.50.11355,PhysRevB.78.075109}. Hence, the Hamiltonian matrix, Eq.~(\ref{eq:ham-matrix-grid}), can be a sparse matrix in the real-space grid representation, and it can be efficiently treated in the time-propagation of quantum systems \cite{SASato2014jasse}.

A strong point of the real-space grid representation is the capability to describe complex electron dynamics because a large number of grid points can capture highly complex wavefunctions. Furthermore, by naturally increasing the number of grid points and reducing grid spacing, the quality of the representation can be straightforwardly improved. Hence, the convergence of the numerical results can be easily checked. Although the real-space grid representation provides large merit in the description of complex electron dynamics, it requires relatively high computational costs due to the large degree of freedom. Especially when the dynamics of (semi) core levels play an important role, the real-space grid representation becomes infeasible due to the huge computational costs since very fine grid spacing is required. To describe such electron dynamics, we employ an alternative numerical method described below.

\subsection{Basis expansion \label{subsubsec:basis}}

Next, we revisit the basis expansion approach for electron dynamics in solids \cite{PhysRevB.89.224305}. Although the real-space grid representation offers a comprehensive description of electron dynamics in solids, it tends to require higher computational costs. To reduce the computational cost of the electron dynamics simulation, we developed an orbital basis expansion approach. In this approach, we expand the time-dependent Kohn-Sham Bloch orbital with eigenstates of the field-free Kohn-Sham Hamiltonian. These eigenstates are defined by
\begin{align}
\hat h_{\bm k}(t=0)u^S_{b\bm k}(\bm r) = \epsilon_{b\bm k}u^S_{b\bm k}(\bm r),
\label{eq:kseq-eig}
\end{align}
where $u^S_{b\bm k}(\bm r)$ are eigenstates, and $\epsilon_{b\bm k}$ are the corresponding eigenvalues. A set of all eigenstates at any $\bm k$-points spans a complete set for periodic functions in the unit cell. Therefore, one may naively consider the following simple expansion with the eigenstates at a single $\bm k$-point
\begin{align}
u_{b\bm k}(\bm r,t)=\sum^{N_{max}}_{b'=1} C^S_{bb' \bm k}(t)u^S_{b'\bm k}(\bm r),
\label{eq:simple-expansion}
\end{align}
where $C^S_{bb'' \bm k}(t)$ are the time-dependent expansion coefficients, and $N_{max}$ eigenstates from the lowest eigenstate are included. Since $\left \{u^S_{b\bm k}(\bm r) \right \}$ is a complete basis set, the expansion of Eq.~(\ref{eq:simple-expansion}) provides the exact description of the time-dependent Kohn-Sham orbitals in the large basis limit $(N_{max}\rightarrow\infty)$. However, in the previous work \cite{PhysRevB.89.224305}, it has been demonstrated that the naive expansion with the eigenstates at a single $\bm k$-point shows very slow convergence for the description of electron dynamics. The origin of the slow convergence of Eq.~(\ref{eq:simple-expansion}) is a poor description of the light-induced intraband transitions. In the single-particle picture, laser fields may induce two kinds of electronic transitions in solids: one is the interband transition, which is the transition between different bands. The other is the intraband transition, which is the transition within the same band and is described by the Bloch momentum shift with the applied laser field as $u^S_{b\bm k}(\bm r)\rightarrow u^S_{b,\bm k+\bm A(t)}(\bm r)$. As seen from the expression of Eq.~(\ref{eq:simple-expansion}), the simple expansion may naturally describe the interband transitions as it contains states in different bands. In contrast, the intraband transitions cannot be directly described in the expansion, and the description of the intraband transitions requires a linear combination of many eigenstates at a single $\bm k$-point, resulting in the slow convergence of Eq.~(\ref{eq:simple-expansion}). In order to achieve faster convergence, we developed yet another expansion by directly including the momentum shifted eigenstates as
\begin{align}
u_{b\bm k}(\bm r,t)&=\sum^{N_{max}}_{b'=1} C^S_{bb' \bm k}(t)u^S_{b'\bm k}(\bm r)
+\sum^{N_{shift}}_{n=1} \sum^{N_{n,max}}_{b'=1} C_{b \bm k b' \Delta \bm k_n}(t)u^S_{b'\bm k+\Delta \bm k_n}(\bm r),
\label{eq:sato-expansion}
\end{align}
where $N_{shift}$ is the number of shifted $\bm k$-points, $\Delta \bm k_n$ is the size of the shift, and $C_{b \bm k b' \Delta \bm k_n}(t)$ are the expansion coefficients. The size of the $\bm k$-shift, $\Delta \bm k_n$, should be chosen such that the intraband transition induced by a field $\bm A(t)$ is suitably described. Since the first term in Eq.~(\ref{eq:sato-expansion}) directly describes the interband transition and the second term describes the intraband transition, the light-induced transitions can be efficiently described. Indeed, the fast convergence and the reduction of the computational cost with the novel expansion, Eq.~(\ref{eq:sato-expansion}), has been demonstrated in the previous work \cite{PhysRevB.89.224305}. The novel basis expansion approach can be used as an alternative computational scheme to simulate electron dynamics in solids when the real-space grid representation is infeasible due to the large computational cost.

\section{Real-time electron dynamics simulations \label{sec:rt-dynamcis}}

Here we review the first-principles analysis of light-induced electron dynamics in solids and demonstrate the basic workflow of its application, by revisiting the recent attosecond experiments \cite{Schlaepfer2018,Volkov2019}. First, the basic linear response calculation in the time domain is reviewed. Then, we further review the extension of the linear response calculation to investigate nonlinear responses of matter. Furthermore, we describe the microscopic analysis of the simulation results on the basis of the microscopic decomposition, providing physics insight into the light-induced ultrafast phenomena.

\subsection{Attosecond electron dynamics in transition metals, $Ti$ \label{subsec:atto-Ti}}

In the previous work \cite{Volkov2019}, we applied the first-principles analysis to the attosecond transient absorption spectroscopy for Titanium and found that an intense femtosecond infrared (IR) laser pulse can change the optical property of transition metals in the sub-femtosecond timescale. On the basis of the microscopic analysis, we clarified that the change of photo-absorption originates from the modification of microscopic screening properties through the light-induced electron localization. Here, we review the first-principles simulations of light-induced electron dynamics in solids by demonstrating the basic workflow of the \textit{ab-initio} investigation of the attosecond transient spectroscopy.

\subsubsection{Linear response calculations in the time domain \label{subsubsec:linear-response}}

To understand dynamical properties of matter, it is important to have solid understanding of the static properties first. Here, we review the linear response calculations to evaluate the dielectric function of solids with the real-time electron dynamics simulation.

In the real-time electron dynamics simulation, the time-propagation of Kohn-Sham orbitals can be evaluated as a sequence of small time-step propagation as
\begin{align}
u_{b\bm k}(\bm r, t+\Delta t) \approx \exp \left [-i \Delta t 
\hat h_{KS,\bm k}\left (t+\frac{\Delta t}{2} \right )
\right ]u_{b\bm k}(\bm r,t),
\label{eq:time-propagation}
\end{align}
where $\Delta t$ is the time step. Here, for the small time-step propagation, the Hamiltonian at $t+\Delta t/2$ is used; this is known as the midpoint rule \cite{doi:10.1063/1.1774980}. To practically operate the exponential operator in Eq.~(\ref{eq:time-propagation}), we employ the fourth order Taylor expansion \cite{PhysRevB.54.4484} as
\begin{align}
u_{b\bm k}(\bm r, t+\Delta t) \approx \sum^4_{n=0} \frac{1}{n!} \left (
-i \Delta t \hat h_{KS,\bm k}\left (t+\frac{\Delta t}{2} \right )
\right )^n u_{b\bm k}(\bm r,t).
\label{eq:time-propagation-Taylor}
\end{align}
Here, the time-propagation is described only by the matrix operation to the state vectors. Thus, the time-propagation can be straightforwardly performed in both representations in Sec.~\ref{sec:numerics}, the real-space grid representation or the basis expansion. In this review, we employ the real-space grid representation to describe the electron dynamics in Titanium. Here, the unit cell containing two Titanium atoms are discretized into $22\times 22\times 35$ grid points. Likewise, the first Brillouin zone is discretized into $20\times 20 \times 20$ $\bm k$-points. As the exchange-correlation potential, the local density approximation is chosen \cite{PhysRevB.45.13244}. Hereafter, the same parameterization will be used for the electron dynamics in Titanium. To perform the electron dynamics simulation practically, we employed Octopus code \cite{doi:10.1063/1.5142502}.

To evaluate optical properties of Titanium, we first compute electron dynamics in the unit cell under an impulsive distortion, $\bm E(t)=E_0 \bm e_z \delta(t)$, where $E_0$ is the strength of the distortion, and $\bm e_z$ is the unit vector along the $z$-axis, which is set to the $c$-axis of Titanium. The corresponding vector potential is given by
\begin{align}
\bm A(t) = - E_{0,z} \bm e_z \Theta(t),
\label{eq:vecpot-delta}
\end{align}
where $\Theta(t)$ is the Heaviside step function. Here, the field strength $E_0$ is set to $5\times 10^{-3}$ a.u., which is so weak that the induced response is a linear response.

By solving the time-dependent Kohn-Sham equation, Eq.~(\ref{eq:tdks-velocity-bloch}), with the vector potential of Eq.~(\ref{eq:vecpot-delta}), we evaluate the electron dynamics induced by the weak distortion. On the basis of Eq.~(\ref{eq:current-density}), the induced current can be evaluated. When the applied field is weak, the induced current and the field have the following linear relation
\begin{align}
J_{\alpha}(t) = \sum_{\beta=\{x,y,z\}}\int^t_{-\infty} dt' \sigma_{\alpha \beta}(t-t')E_{\beta}(t'),
\label{eq:linear-response-td}
\end{align}
where $J_{\alpha}(t)$ is the $\alpha$-component of $\bm J(t)$, $E_{\alpha}(t)$ is the $\alpha$-component of $\bm E(t)$, and $\sigma_{\alpha \beta}(t)$ is the optical conductivity in the time domain. By applying Fourier transform to Eq.~(\ref{eq:linear-response-td}), one can rewrite Eq.~(\ref{eq:linear-response-td}) as
\begin{align}
J_{\alpha}(\omega) = \sum_{\beta =\{x,y,z\}} \sigma_{\alpha \beta}(\omega)E_{\beta}(\omega),
\label{eq:linear-response-fd}
\end{align}
where $J_{\alpha}(\omega)$, $E_{\alpha}(\omega)$, and $\sigma_{\alpha \beta}(\omega)$ are the Fourier transform of $J_{\alpha}(t)$, $E_{\alpha}(t)$, and $\sigma_{\alpha \beta}(\omega)$, respectively. Furthermore, one can evaluate the dielectric function as
\begin{align}
\epsilon_{\alpha \beta} = \delta_{\alpha \beta} + \frac{4\pi i}{\omega}\sigma_{\alpha \beta}(\omega).
\label{eq:dielectric-function}
\end{align}

\begin{figure}[htb]
\centering\includegraphics[width=1.00\linewidth]{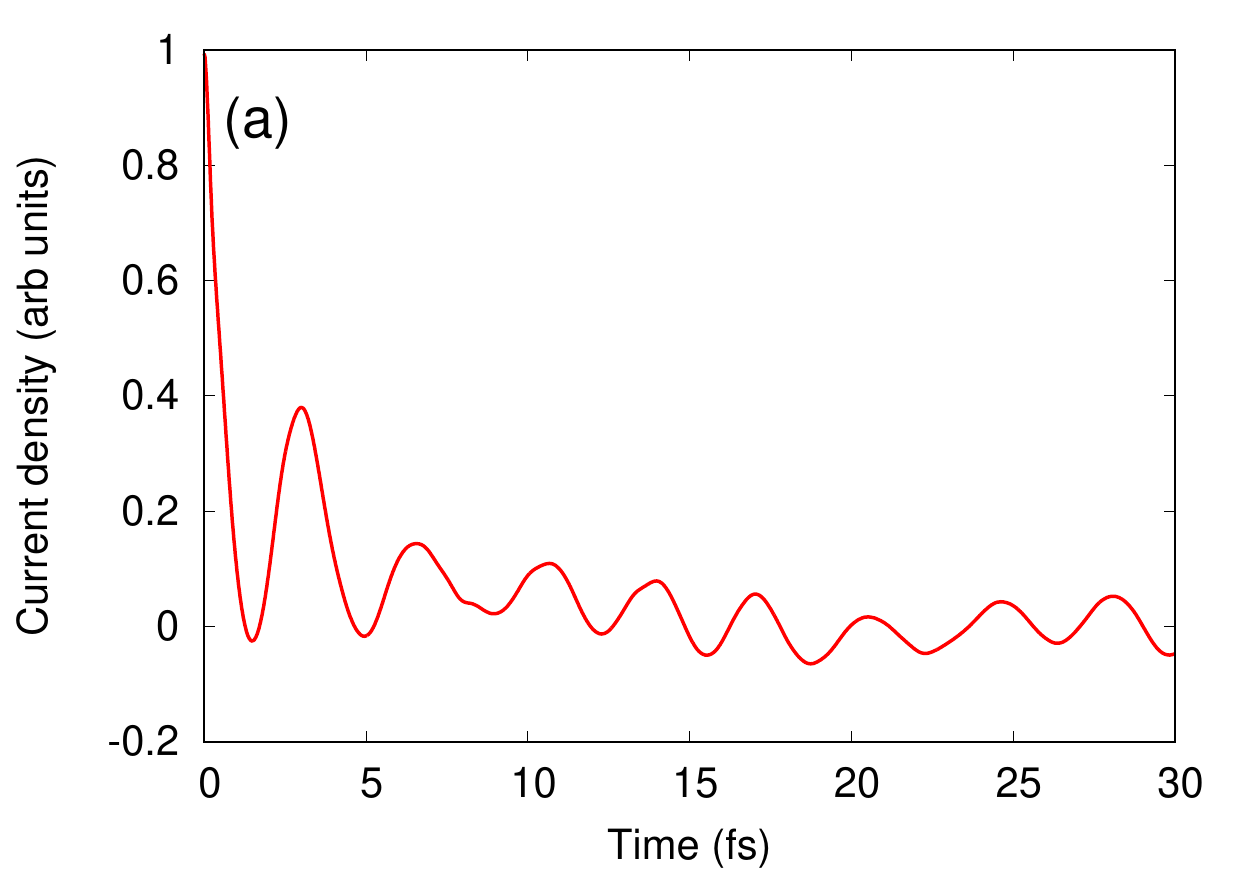}
\centering\includegraphics[width=1.00\linewidth]{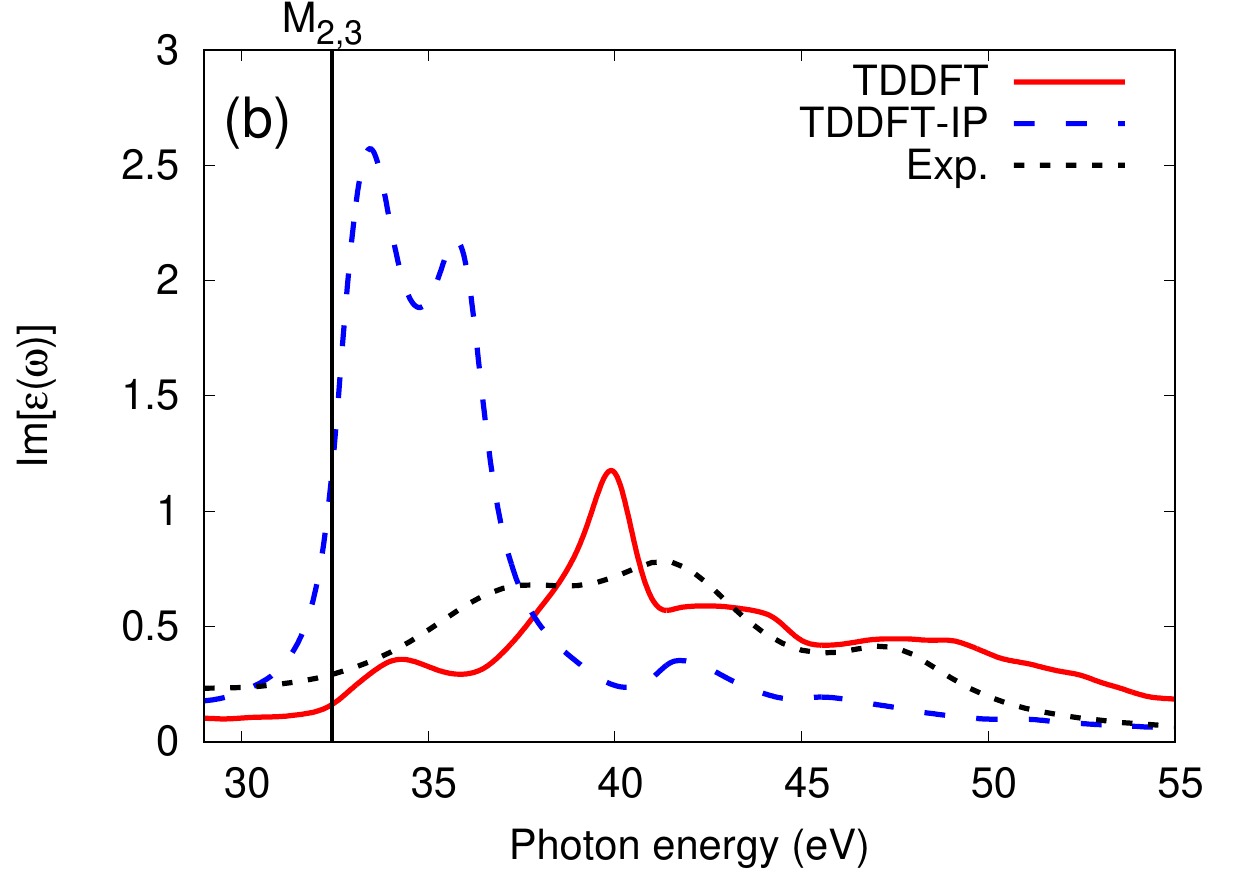}
\caption{\label{fig:current_Ti}
Linear response calculation for bulk Titanium. (a) the electric current induced by the impulsive distortion at $t=0$. (b) the imaginary part of the dielectric function of Titanium. The results of the TDDFT simulation (red-solid line), the independent particle approximation (blue-dashed line), and the experiment (black-dotted line) \cite{doi:10.1063/1.3243762} are shown. The computed absorption edge is indicated by the black-solid line.}
\end{figure}

In Fig.~\ref{fig:current_Ti}~(a), the $z$-component of the current induced by the impulsive distortion, Eq.~(\ref{eq:vecpot-delta}), is shown. The current is suddenly induced at $t=0$, and it shows the oscillation. With the induced current, one can evaluate the diagonal element of the optical conductivity as $\sigma_{zz}(\omega) = J_z(\omega)/E_{0,z}$ and the dielectric function as $\epsilon_{zz}(\omega)=1+4\pi i \sigma_{zz}/\omega$. By repeating the same procedure with different impulsive distortion along another direction, one can further evaluate the other diagonal parts of the dielectric function, such as $\epsilon_{xx}(\omega)$ and $\epsilon_{yy}(\omega)$. Further averaging the different components of the dielectric function, one can evaluate the orientation averaged dielectric function as
\begin{align}
\epsilon(\omega) = \frac{
\epsilon_{xx}(\omega)+\epsilon_{yy}(\omega)+\epsilon_{zz}(\omega)
}{3}.
\end{align}
The orientation averaged dielectric function can be used to evaluate the dielectric function of polycrystalline samples.

Figure~\ref{fig:current_Ti}~(b) shows the imaginary part of the dielectric function of Titanium. The red-solid line shows the computed dielectric function $\epsilon(\omega)$ by TDDFT with the adiabatic local density approximation \cite{PhysRevB.45.13244}, and the black-dotted line shows the experimental result \cite{doi:10.1063/1.3243762}. The $M_{2,3}$ absorption edge is indicated by the black-solid line. We further evaluated the dielectric function with the independent particle (IP) approximation, where the time-dependence of the Hartree and exchange-correlation potentials, $v_H(\bm r, t)$ and $v_{xc}(\bm r, t)$, are ignored, and these potentials are frozen at $t=0$. The result of the IP approximation is shown as the blue-dashed line in Fig.~\ref{fig:current_Ti}~(b). One sees that the result of the TDDFT (red-solid line) shows the fair agreement with the experimental data (black-dotted line) while that of the IP approximation (blue-dashed line) shows a significantly different behavior: the results of the IP approximation shows a strong absorption just above the $M_{2,3}$ edge while that of the TDDFT shows a large spectral weight around $40$~eV. The blue shift of the spectral weight reflects the collective excitation in the TDDFT calculation via the microscopic electronic screening effect, which is completely ignored in the IP approximation. This screening effect is also known as the local field effect, and it tends to be strong when the initial and final states of the corresponding transitions are spatially localized \cite{PhysRevB.60.R16251}.

\subsubsection{Numerical pump-probe experiments \label{subsubsec:pump-probe}}

In the previous section, Sec.~\ref{subsubsec:linear-response}, we evaluated the optical properties of solids with the electric current induced by a weak perturbation. Here, we describe an extension of the linear response calculation to nonlinear regime, by mimicking experimental pump-probe techniques by the first-principles electron dynamics simulation. We shall call it \textit{numerical pump-probe experiment} \cite{PhysRevB.89.064304}.

The numerical pump-probe experiment is a similar approach to the linear response calculation in Sec.~\ref{subsubsec:linear-response}. In the numerical pump-probe experiment, the electron dynamics induced by pump and probe electric fields, $\bm E_{pump}(t)$ and $\bm E_{probe}(t)$, is simulated, and the induced current is evaluated. We shall call the evaluated current \textit{pump-probe current} $\bm J_{pump-probe}(t)$, which is induced by the pump and probe electric fields. Then, we simulate the electron dynamics induced solely by the pump pulse $\bm E_{pump}(t)$ and evaluate the induced current. We shall call the evaluated current \textit{pump current} $\bm J_{pump}(t)$, which is induced solely by $\bm E_{pump}(t)$. To extract the current induced by the probe field $\bm E_{probe}(t)$ under the presence of the pump field $\bm E_{pump}(t)$, we define the \textit{probe current} as the difference of the pump-probe current and the pump current:
\begin{align}
\bm J_{probe}(t)\equiv \bm J_{pump-probe}(t)-\bm J_{pump}(t).
\end{align}
One can evaluate transient optical properties of solids under the presence of pump fields by analyzing the probe current $\bm J_{probe}(t)$ and the probe field $\bm E_{probe}(t)$. For example, the transient optical conductivity can be evaluated as
\begin{align}
\sigma^{ex}_{\alpha \beta}(\omega) = \frac{J_{probe,\alpha}(\omega)}{E_{probe,\beta}(\omega)},
\label{eq:pump-probe-sigma}
\end{align}
where $E_{probe,\alpha}(\omega)$ and $J_{probe,\alpha}(\omega)$ are the $\alpha$-components of the Fourier transform of the probe field $\bm E_{probe}(\omega)$ and the probe current $\bm J_{probe}(\omega)$, respectively. Furthermore, the transient dielectric function can be evaluated with the transient optical conductivity as
\begin{align}
\epsilon^{ex}_{\alpha \beta} = \delta_{\alpha \beta} + \frac{4\pi i}{\omega}\sigma^{ex}_{\alpha \beta}(\omega).
\end{align}

If numerical pump-probe experiments are performed with time-delay $T_{delay}$ between the pump and probe pulses, one can evaluate the transient dielectric function as a function of the time-delay $T_{delay}$. In the previous work \cite{Volkov2019}, we studied the transient optical property of Titanium under the presence of the femtosecond IR laser pulse. In the simulation, we employed the following vector potential for the pump field:
\begin{align}
\bm A_{pump}(t) = -\frac{E_{pump,0}}{\omega_{pump}}\bm e_z \sin \left (\omega t \right)
\cos^2 \left (\frac{\pi}{T_{pump}} t \right )
\label{eq:pump-Ti}
\end{align}
in the duration $-T_{pump}/2<t<T_{pump}/2$ and zero outside. Here, the peak field strength $E_{pump,0}$ is set to $9.7$~MV/cm, the mean frequency $\omega_{pump}$ is set to $1.55$~eV$/\hbar$, and the pulse duration $T_{pump}$ is set to $20$~fs.

For the probe field, we employ the following impulsive distortion
\begin{align}
\bm A_{probe}(t) = -E_{pump,0} \bm e_z  \Theta \left (t -T_{delay}  \right ),
\end{align}
where the strength of the impulse $E_{probe, 0}$ is set to $10^{-3}$~a.u.. Here, $T_{delay}$ is the time delay between the pump and probe pulses. We repeated the pump-probe simulation by changing the time-delay $T_{delay}$ and evaluated the transient dielectric function $\epsilon(\omega, T_{delay})$.

\begin{figure}[htb]
\centering\includegraphics[width=1.00\linewidth]{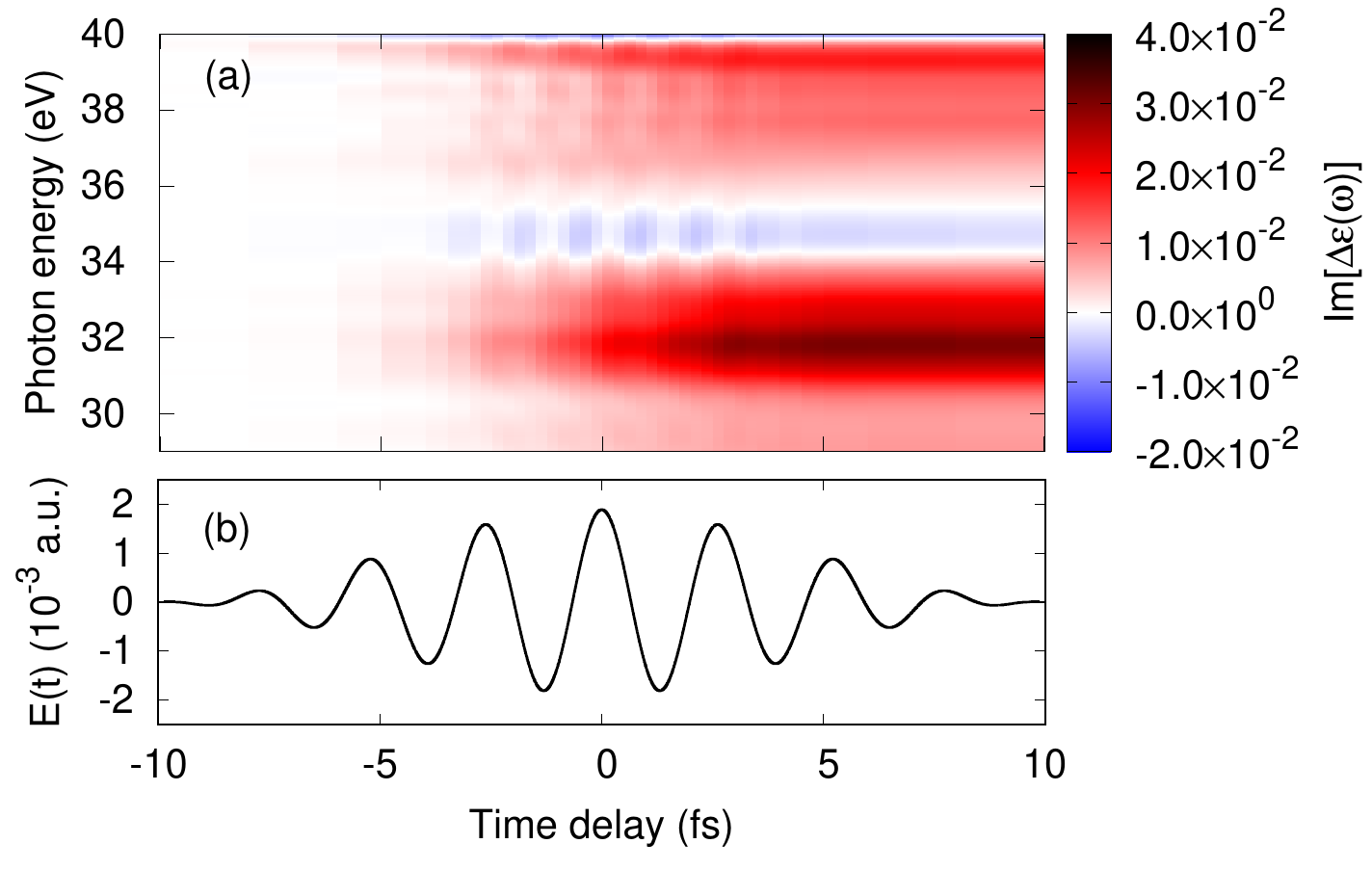}
\caption{\label{fig:delta_eps_im_tr}
(a) The imaginary part of the transient dielectric function $\Delta \epsilon(\omega, T_{delay})$ of Titanium computed by the numerical pump-probe experiments \cite{Volkov2019}. (b) The time profile of the pump electric field defined by Eq.~(\ref{eq:pump-Ti}).
}
\end{figure}

Figure~\ref{fig:delta_eps_im_tr}~(a) shows the imaginary part of the change of the transient dielectric function, $\Delta \epsilon(\omega, T_{delay}) = \epsilon(\omega, T_{delay})- \epsilon(\omega, T_{delay}=-\infty)$, under the presence of the pump pulse. The time profile of the pump electric field is shown in Fig.~\ref{fig:delta_eps_im_tr}~(b). As seen from the figures, the photo-absorption around $32$~eV, which corresponds to the $M_{2,3}$ edge, is enhanced by the pump field irradiation. The similar enhancement of the photo-absorption has been observed in the experiment \cite{Volkov2019}. Hence it has been demonstrated that the first-principles pump-probe simulations can fairly reproduce the results of the attosecond transient absorption spectroscopy.

\subsubsection{Linear response calculation with finite electron-temperature \label{subsubsec:linear-response-FT}}

The above numerical pump-probe experiments opened a novel avenue to explore light-induced nonequilibrium phase of matter from first-principles. In real systems, the light-induced nonequilibrium states then show the relaxation dynamics. Shortly after the laser irradiation, the excited electronic system is thermalized and forms a finite electron-temperature state (hot electron state), where the electron temperature is much higher than the lattice (phonon) temperature. To study optical properties of such quasi-equilibrium states, we developed the linear response calculation with the finite electron-temperature \cite{PhysRevB.90.174303}. To perform the finite electron-temperature calculation, we extend the linear response calculation introduced in Sec.~\ref{subsubsec:linear-response} by modifying the occupation factor with the Fermi-Dirac distribution,
\begin{align}
f^{T_e}_{b\bm k} = \frac{1}{e^{(\epsilon^{T_e}_{b\bm k}-\mu)/k_B T_e}+1},
\end{align}
where $T_e$ is the electron temperature, $\mu$ is the chemical potential, and $\epsilon^{T_e}_{b\bm k}$ is the single-particle energy. In the finite electron-temperature calculation, physical quantities are evaluated with the modified occupation factor. For example, the electron density is given by
\begin{align}
\rho^{T_e}(\bm r,t) = \sum_{b} \int_{BZ} d\bm k f^{T_e}_{b\bm k} \int_{\Omega} d \bm r
\left |u^{T_e}_{b\bm k}(\bm r, t) \right |^2.
\label{eq:density-finite-temperature}
\end{align}
Likewise the current density is given by
\begin{align}
\bm J^{T_e}(t) = -\frac{1}{\Omega} \sum_{b} \int_{BZ} d\bm k f^T_{b\bm k} \int_{\Omega} d \bm r
u^{T_e,*}_{b\bm k}(\bm r, t) \bm \pi_{\bm k}(t) u^{T_e}_{b\bm k}(\bm r, t).
\label{eq:currentfinite-temperature}
\end{align}
Note that the Kohn-Sham potential, Eq.~(\ref{eq:ks-pot}), is evaluated with the electron density at finite temperature $\rho^{T_e}(\bm r,t)$. Hence the initial condition of the time-dependent Kohn-Sham equation should be self-consistently prepared with the occupation factors $f^{T_e}_{b\bm k}$, which are defined by the eigenvalues of the Kohn-Sham Hamiltonian with the ground electron density at the finite electron temperature,
\begin{align}
\rho^{T_e}_{gs}(\bm r) = \sum_{b} \int_{BZ} d\bm k f^{T_e}_{b\bm k} \int_{\Omega} d \bm r
\left |u^{T_e}_{b\bm k}(\bm r, t=-\infty) \right |^2.
\label{eq:density-gs-finite-temp}
\end{align}
Besides the occupation factor $f^{T_e}_{b\bm k}$, we follow the same procedure as the linear response calculation in Sec.~\ref{subsubsec:linear-response}.

\begin{figure}[htb]
\centering\includegraphics[width=1.00\linewidth]{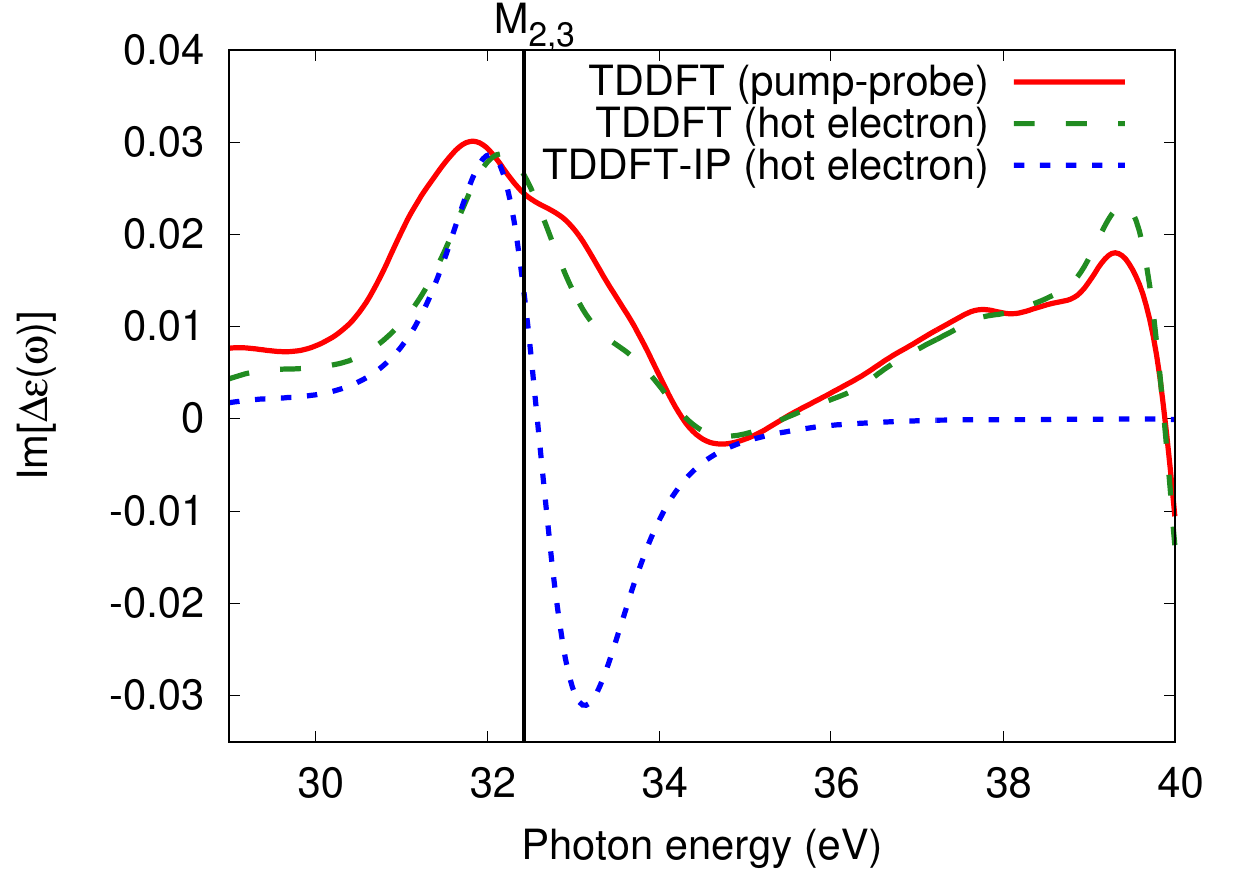}
\caption{\label{delta_eps}
The change of the imaginary part of the dielectric function of Titanium by the electronic excitation. The result of the pump-probe simulation with TDDFT (red-solid line) and that of the finite electron-temperature calculation with TDDFT (green-dashed line) are shown. The result of the finite electron-temperature calculation with the IP approximation (blue-dotted line) is also shown, but it is scaled by a factor of $0.2$.}
\end{figure}

To investigate the optical property of the optically-pumped Titanium, we evaluate the dielectric function with the finite electron-temperature linear response calculation by setting the electron temperature to $T_e=0.315$~eV/$k_B$. Figure~\ref{delta_eps} shows the change of the imaginary part of the dielectric function by the electron temperature increase. The green-dashed line shows the result of the finite electron-temperature calculation with $T_e=0.315$~eV/$k_B$, while the red-solid line shows the result of the numerical pump-probe simulation in Sec.~\ref{subsubsec:pump-probe} at the large delay $T_{delay}=10$~fs. As seen from the figure, the finite electron-temperature calculation and the numerical pump-probe experiment provide the qualitatively same results, indicating that the non-equilibrium and quasi-equilibrium distributions show the consistent behavior. Therefore, the modification of the optical property does not significantly depend on the nature of excited electron distribution, nonequilibrium or quasi-equilibrium.

In Fig.~\ref{delta_eps}, the blue-dotted line shows the result of the finite electron-temperature calculation with the independent particle (IP) approximation, where the time-dependence of the Hartree and exchange-correlation potentials are ignored. Although the results of the full TDDFT calculations (red-solid and green-dashed lines) show the peak structure around $32$~eV, the result of the IP approximation does not show the peak structure but the dispersive structure, where the absorption is enhanced below the $M_{2,3}$ edge while it is suppressed above the edge. Therefore, one can clearly conclude that the peak structure around $32$~eV originates from a many-body effect.

The origin of the dispersive structure in the result of the IP approximation can be straightforwardly understood by the state-filling effect: The $M_{2,3}$ edge corresponds to the transition from the semi-core states ($3p$-states of Ti) to the Fermi level. When the electronic system is excited by light or temperature, electrons below the Fermi level are transferred above the Fermi level. As a result, holes are created below the Fermi level, and electrons are populated above the Fermi level. Due to the formation of holes, the photo-absorption below the $M_{2,3}$ edge is enhanced. On the other hand, due to the population of electrons above the Fermi level, the photo-absorption is suppressed by the Pauli blocking above the $M_{2,3}$ edge.

\subsubsection{Microscopic decomposition of transient optical property and real-time real-space electron dynamics in solids \label{subsubsec:microscopic-decomposition}}

To clarify the origin of the peak structure around the $M_{2,3}$ edge in Fig.~\ref{delta_eps}, we developed the microscopic decomposition of the transient optical properties into several components \cite{Volkov2019}. Here, we describe the details of the decomposition. For this purpose, we revisit the basic equations for the linear response calculations. The electron dynamics of the Kohn-Sham system at the finite electron temperature $T_e$ under a weak perturbation is described by the following time-dependent Kohn-Sham equation,
\begin{align}
i \frac{\partial}{\partial t} u^{T_e}_{b\bm k}(\bm r, t) = \hat h^{T_e}_{KS,\bm k}(t) u^{T_e}_{b\bm k}(\bm r, t),
\end{align}
with the Kohn-Sham Hamiltonian $\hat h^{T_e}_{KS,\bm k}(t)$ at the finite electron temperature. According to Eq.~(\ref{eq:tdks-velocity-bloch-ham}), it is given by
\begin{align}
\hat h^{T_e}_{KS,\bm k}(t) = & 
\frac{1}{2}\left (\bm p + \bm k + \bm A(t) \right )^2 + e^{-i\left (\bm A(t) + \bm k \right )\cdot \bm r} v_{ion}(\bm r, t)e^{i \left ( \bm A(t) + \bm k \right )\cdot \bm r}
 \nonumber \\
&+ v_{Hxc}\left [\rho^{T_e}(\bm r,t) \right](\bm r, t),
\label{eq:tdks-finite-temperature}
\end{align}
where $v_{Hxc}\left [\rho^{T_e}(\bm r,t) \right](\bm r, t)$ is the sum of th Hartree and exchange-correlation potentials. Here we define the induced density as
\begin{align}
\delta \rho^{T_e}(\bm r,t) = \rho^{T_e}(\bm r, t) - \rho^{T_e}_{gs}(\bm r).
\label{eq:delta-density-finite-temp}
\end{align}
Assuming that the perturbation is weak and the induced density $\delta \rho^{T_e}(\bm r,t)$ is small enough, the Kohn-Sham Hamiltonian can be expanded as
\begin{align}
\hat h^{T_e}_{KS,\bm k}(t) = & 
\frac{1}{2}\left (\bm p + \bm k + \bm A(t) \right )^2 + e^{-i\left (\bm A(t) + \bm k \right )\cdot \bm r} v_{ion}(\bm r, t)e^{i \left ( \bm A(t) + \bm k \right )\cdot \bm r}
 \nonumber \\
&+ v_{Hxc}[\rho^{T_e}_{gs}(\bm r)](\bm r) + \int dr'dt' f_{Hxc}(\bm r, \bm r',t,t')\delta \rho^T(\bm r',t'),
\label{eq:tdks-linear}
\end{align}
where $f_{Hxc}(\bm r, \bm r',t,t')$ is the Hartree-exchange-correlation kernel defined as the functional derivative of the Hartree-exchange-correlation potential with respect to the density, $f_{Hxc}(\bm r, \bm r',t,t') = \delta v_{Hxc}[\rho(\bm r, t)]/\delta \rho(\bm r', t')$.

The electron temperature $T_e$ affects the linear response through the three equations; Eq.~(\ref{eq:currentfinite-temperature}), Eq.~(\ref{eq:density-gs-finite-temp}) and Eq.~(\ref{eq:delta-density-finite-temp}). Hence, one can decompose the temperature effect into the following three components: one is the contribution from the ground state density $\rho^{T_e}_{gs}(\bm r)$ in Eq.~(\ref{eq:density-gs-finite-temp}). We shall call it \textit{band-structure renormalization} since it results in the modification of the electronic structure though the ground-state Hartree-exchange-correlation potential $v_{Hxc}[\rho^T_{gs}(\bm r)](\bm r)$. The second one is the contribution from the induced density $\delta \rho^{T_e}(\bm r,t)$ in Eq.~(\ref{eq:delta-density-finite-temp}). We shall call it \textit{modification of the local field effect} because the main contribution is the modification of the microscopic screening effect though the Hartree-exchange-correlation kernel $f_{Hxc}(\bm r, \bm r',t,t')$ in Eq.~(\ref{eq:tdks-linear}). The last one is the contribution from the occupation factor in the current evaluation in Eq.~(\ref{eq:currentfinite-temperature}). We shall call it \textit{modification of the state-filing effect}.

\begin{figure}[htb]
\centering\includegraphics[width=1.00\linewidth]{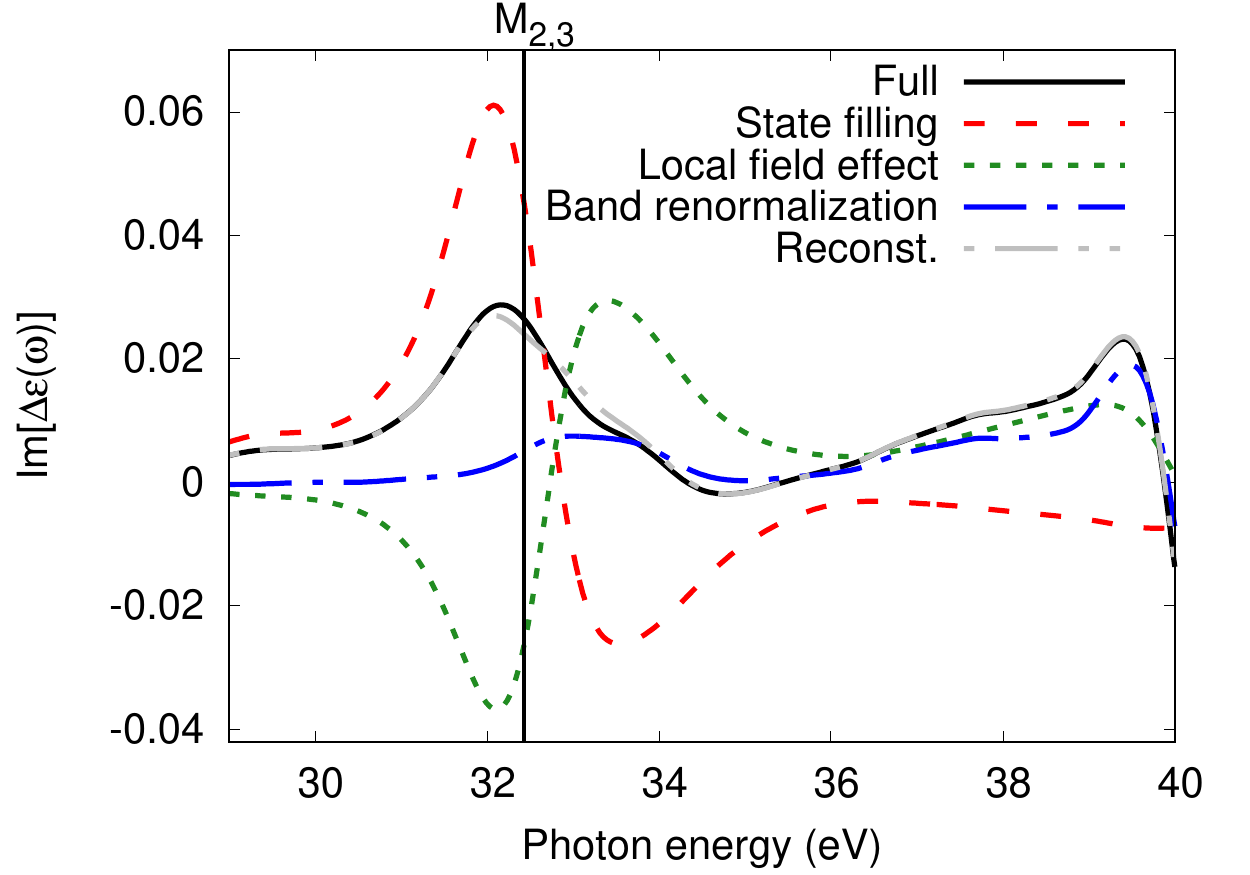}
\caption{\label{fig:delta_eps_decomp}
The change of the dielectric function $\Delta \epsilon^T(\omega)$. The result of the full TDDFT calculation with the finite electron temperature is shown as the black-solid line. The contributions from the state-filling (red-dashed line), the local field modification (green-dotted line), and the band renormalization (blue-dash-dot line) are shown. The reconstructed signal from the three decomposed contributions is shown as the gray dash-dot-dot line.
}
\end{figure}

One can evaluate each contribution of (i) the band structure renormalization, (ii) modification of the local field effect, and (iii) the state-filling effect, by setting the finite temperature of $T_e=0.315$~eV/$k_B$ to one of the three equations, Eq.~(\ref{eq:currentfinite-temperature}), Eq.~(\ref{eq:density-gs-finite-temp}) and Eq.~(\ref{eq:delta-density-finite-temp}),  and by setting the low temperature $T_e=0.01$~eV/$k_B$ to the other equations. Figure~\ref{fig:delta_eps_decomp} shows the change of dielectric function. The black-solid line shows the full TDDFT result, which is also shown as the green-dashed line in Fig.~\ref{fig:delta_eps_im_tr}. In Fig.~\ref{fig:delta_eps_decomp}, the red-dashed line shows the contribution from the state-filling with Eq.~(\ref{eq:currentfinite-temperature}), the green-dotted line shows the contribution from the local field effect with Eq.~(\ref{eq:delta-density-finite-temp}), and the blue-dash-dot line shows the band renormalization contribution with Eq.~(\ref{eq:density-gs-finite-temp}). The reconstructed signal by summing the three decomposed contributions is also shown as the gray dash-dot-dot line, and it fairly reproduces the full signal. Hence, we confirm that the three-components are indeed the decomposed contributions from the full signal.

One sees that the state-filling contribution shows the enhancement of the signal below the $M_{2,3}$ edge and the suppression above the $M_{2,3}$ edge. This is nothing but the Pauli blocking effect, which is discussed to understand the result of the independent particle approximation (blue dotted line) in Fig.~\ref{delta_eps}. In contrast, the modification of the local field (the green-dotted line in Fig.~\ref{fig:delta_eps_decomp}) shows the opposite behavior; the photo-absorption is enhanced above the $M_{2,3}$ edge while it is suppressed below the $M_{2,3}$ edge. As a result of the partial cancellation between the state-filling contribution and the local field effect contribution, the positive peak structure in the transient signal is formed. Therefore, the modification of the microscopic screening effect is a key of the formation of the single peak structure around the $M_{2,3}$ edge.

\begin{figure*}[htb]
\centering\includegraphics[width=0.75\linewidth]{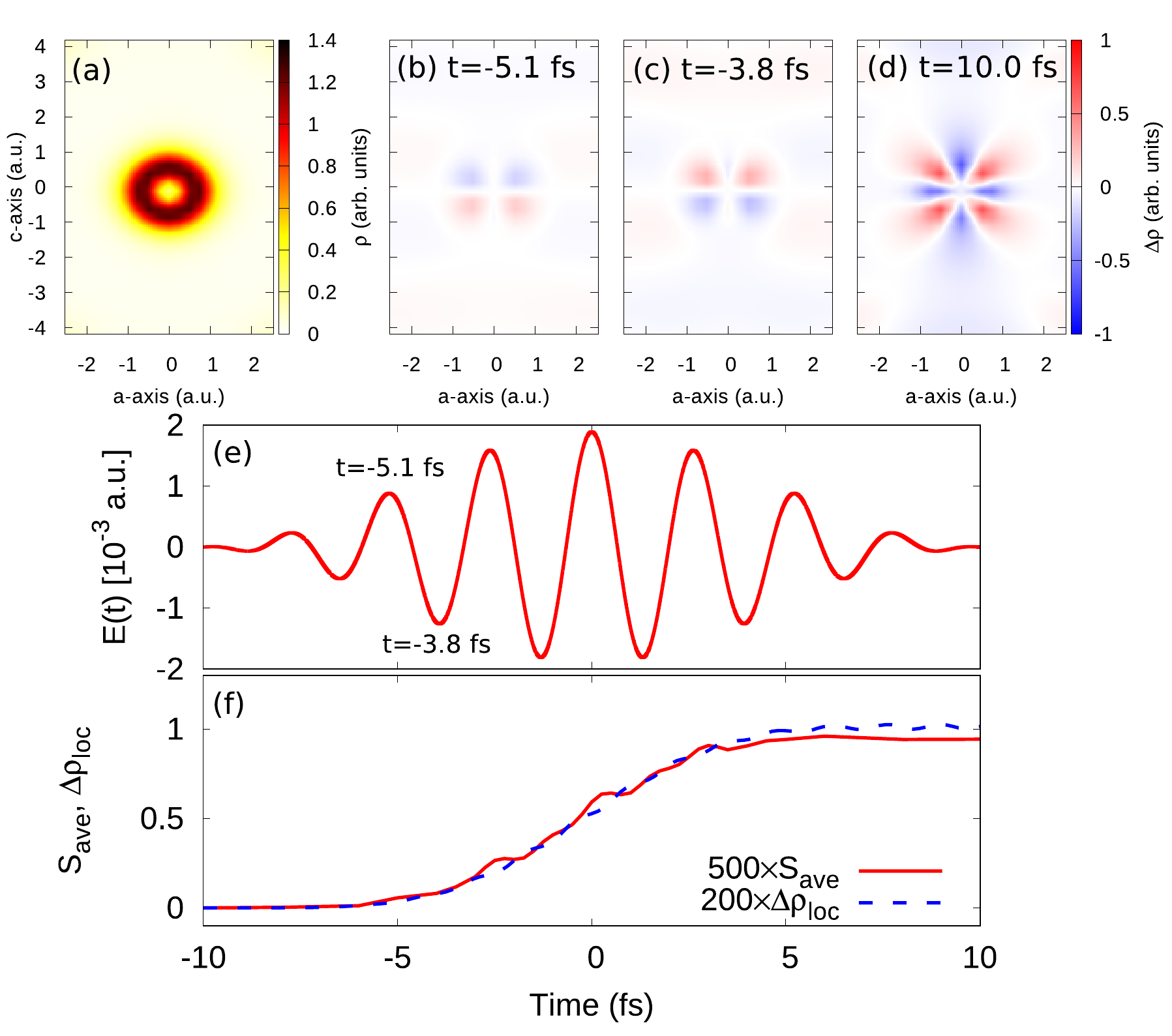}
\caption{\label{fig:Ti_dynamics}
The electron dynamics in Titanium: (a) the ground state electron density. (b-d) the change of the electron density under the pump laser pulse irradiation. (e) the time-profile of the pump laser pulse. (f) the photon-energy averaged transient signal $S_{ave}(T_{delay})$ in Eq.~(\ref{eq:averaged-singal-Ti}), and the localized electron density $\Delta \rho_{loc}(t)$ in Eq.~(\ref{eq:localized-electron-density}) are shown.
}
\end{figure*}

Although the state-filling effect can be clearly understood by the Pauli principle, the microscopic origin of the modification of the local field effect is not clarified yet in the above analysis. To provide further insight into the local field effect modification, we evaluated the real-time real-space electron density dynamics under the pump laser pulse. Figure~\ref{fig:Ti_dynamics}~(a) shows the ground state electron density in Titanium on a plane that consists of Ti ion at the center. One sees that the electron density is localized around the Ti ion. The localized nature of the electron density causes the strong local field effect in the static absorption in Fig.~\ref{fig:current_Ti} \cite{PhysRevB.60.R16251}. Figures~\ref{fig:Ti_dynamics}~(b)-(d) show the snapshots of the change of the electron density from the ground state, $\rho(\bm r,t)-\rho(\bm r,t=-\infty)$. The time profile of the applied electric field is also shown in Fig.~\ref{fig:Ti_dynamics}~(e), which is identical to Fig.~\ref{fig:delta_eps_im_tr}~(b). The earlier part of the laser pulse induces the polarization dynamics, showing the electron density displacement along the opposite direction to the instantaneous field direction (see Figs.~\ref{fig:Ti_dynamics}~(b) and (c)). In contrast, the electron density shows the localization around the Ti ion after the laser irradiation (see Fig.~\ref{fig:Ti_dynamics}~(d)). Thus, we can confirm that the laser excitation induces the electron localization around the Ti ion from first-principles.

To quantify the light-induced electron localization, we evaluate the number of electrons that are localized around the Ti ion by
\begin{align}
\Delta \rho_{loc}(t) = \int d\bm r \left [\rho(\bm r,t)-\rho(\bm r,t=-\infty) \right ]  \mathrm{exp}\left [
-\frac{\left |\bm r - \bm R_{Ti} \right |^2}{2\sigma^2}
\right ],
\label{eq:localized-electron-density}
\end{align}
where $\bm R_{Ti}$ is the position of the titanium ion, and $\sigma$ is the width of the Gaussian window, which is set to $\sigma=1$~a.u. Furthermore, we evaluated the signal intensity of the transient absorption by averaging the transient dielectric function around the $M_{2,3}$ edge as
\begin{align}
S_{ave}(T_{delay}) = \frac{1}{\omega_f - \omega_i} \int^{\omega_f = 31~\mathrm{ev}/\hbar}_{\omega_i = 33~\mathrm{ev}/\hbar} \mathrm{Im} \left [\epsilon(\omega, T_{delay}) \right ].
\label{eq:averaged-singal-Ti}
\end{align}

Figure~\ref{fig:Ti_dynamics}~(f) shows the evaluated signal intensity $S_{ave}(T_{delay})$ and the localized electron density $\Delta \rho_{loc}(T_{delay})$ as functions of the time delay. One can clearly see that the signal intensity is synchronized with the electron localization. This fact clearly indicates that the ultrafast modification of the transient optical property of Titanium in Fig.~\ref{fig:delta_eps_im_tr} originates from the laser-induced ultrafast electron localization via the modification of the local field effect.

\subsection{Attosecond electron dynamics in semiconductors, $GaAs$ \label{subsec:atto-GaAs}}

In the previous section, we reviewed the microscopic decomposition of the transient optical property on the basis of the physical mechanisms: the state-filling, the local-field effect, and the band-structure renormalization. Here, we review yet another microscopic decomposition of the transient optical property, the orbital and band decomposition of dielectric function. For this purpose, we demonstrate the basic workflow of the \textit{ab-initio} analysis by revisiting its application to the recent attosecond transient absorption spectroscopy for GaAs \cite{Schlaepfer2018}. For this analysis, we employed the basis expansion method introduced in Sec.~\ref{subsubsec:basis} with the independent particle approximation, where the time-dependence of the Hartree and exchange-correlation potentials is ignored.

\begin{figure}[htb]
\centering\includegraphics[width=1.00\linewidth]{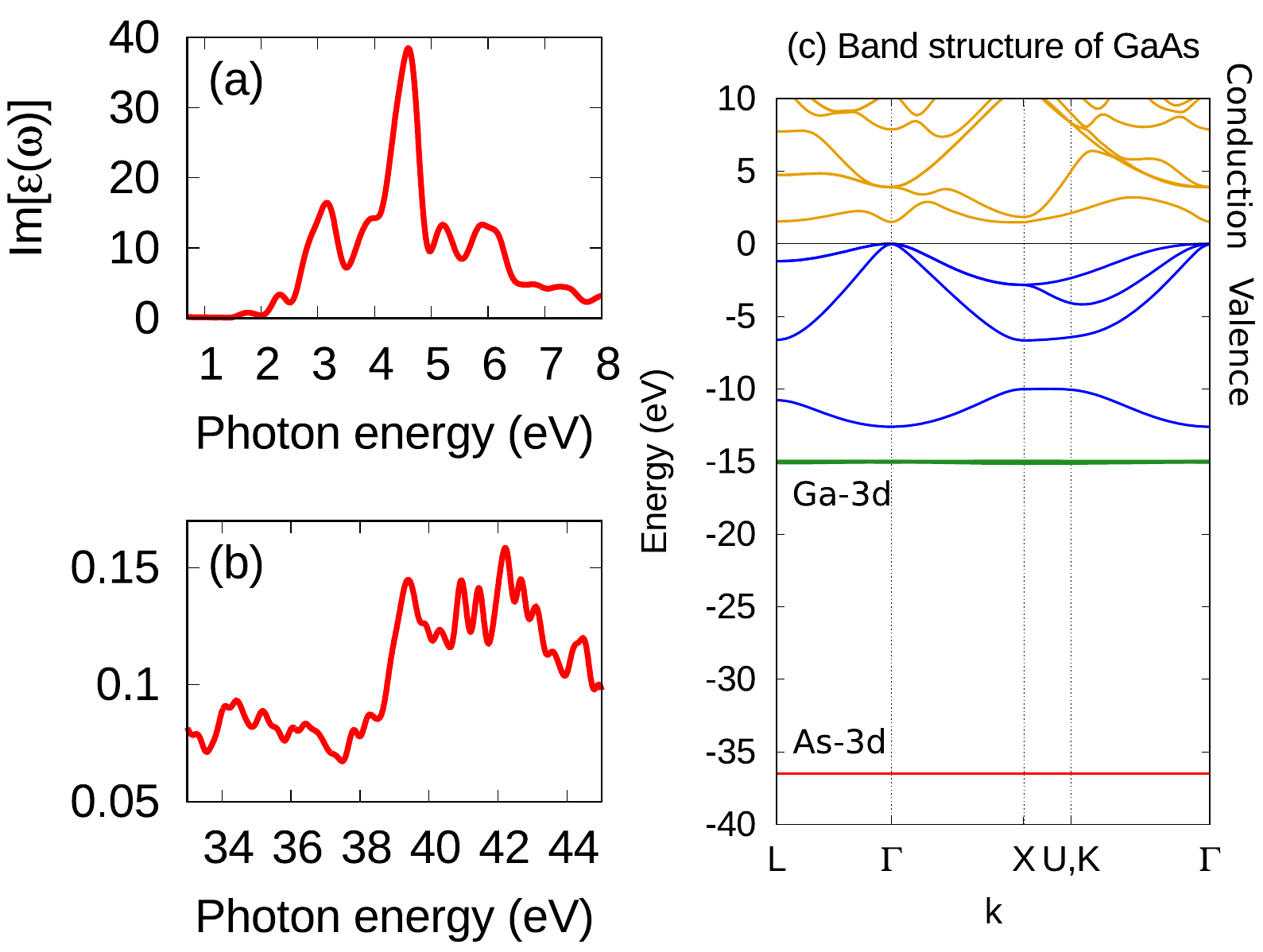}
\caption{\label{fig:GaAs_lin}
The imaginary part of the dielectric function of GaAs in the low energy region (a) and the high energy region (b). (c) the band structure of GaAs.
}
\end{figure}

In the attosecond transient absorption spectroscopy for GaAs \cite{Schlaepfer2018}, the modulation of the photo-absorption related to the As-$3d$ core level has been investigated under the presence of femtosecond IR laser pulses. According to the previous section~\ref{subsec:atto-Ti}, we first analyze the static optical property of GaAs before analyzing transient optical properties. Figures~\ref{fig:GaAs_lin}~(a) and (b) show the imaginary parts of the dielectric function of GaAs in the low and high energy regions, respectively. The computed band-structure is also shown in Fig.~\ref{fig:GaAs_lin}~(c). For these calculations, we employed the modified Becke-Johnson potential \cite{PhysRevLett.102.226401} with the optimized $c$-parameter \cite{PhysRevB.85.155109} to reproduce the band-gap of GaAs. As seen from Fig.~\ref{fig:GaAs_lin}~(b), the clear absorption edge is observed around $38$~eV, which corresponds to the transition from the As-$3d$ core level to the bottom of the conduction band.

Then, we analyze the transient optical property of GaAs by simulating the pump-probe experiment \cite{Schlaepfer2018}. For this purpose, we employed the numerical pump-probe experiments introduced in Sec.~\ref{subsubsec:pump-probe} with the following vector potentials:
\begin{align}
\bm A_{pump}(t) = -\frac{E_{pump,0}}{\omega_{pump}}\bm e_a \sin(\omega_{pump} t)
\cos^2 \left (\frac{\pi}{T_{pump}}t \right )
\label{eq:pump-GaAs}
\end{align}
in the duration $-T_{pump}/2<t<T_{pump}/2$ and zero outside, and
\begin{align}
\bm A_{probe}(t) =& -\frac{E_{probe,0}}{\omega_{probe}}\bm e_a \sin \left (\omega_{probe} \left (t -T_{delay}\right ) \right ) \nonumber \\
&\times \cos^4 \left (\frac{\pi}{T_{probe}} \left (t - T_{delay}\right )  \right )
\end{align}
in the duration $-T_{probe}/2<t-T_{delay}<T_{probe}/2$ and zero outside. Here $\bm e_a$ is the unit vector along the $(110)$-direction of the crystal. The peak field strength of the pump pulse is set to $E_{pump,0}=2.75\times 10^{9}$~V/cm, and that of the probe pulse is set to $E_{probe,0}=2.75 \times 10^8$~V/cm. The mean frequencies of the pump and probe pulses are set to $\omega_{pump}=1.55$~eV/$\hbar$ and $\omega_{probe}=40$~eV/$\hbar$, respectively. The pulse durations of the pump and probe pulses are set to $T_{pump}=14$~fs and $T_{probe}=1$~fs, respectively. We repeated the pump-probe simulation by changing the time-delay $T_{delay}$ and evaluated the transient dielectric function $\epsilon(\omega, T_{delay})$.

\begin{figure}[htb]
\centering\includegraphics[width=1.00\linewidth]{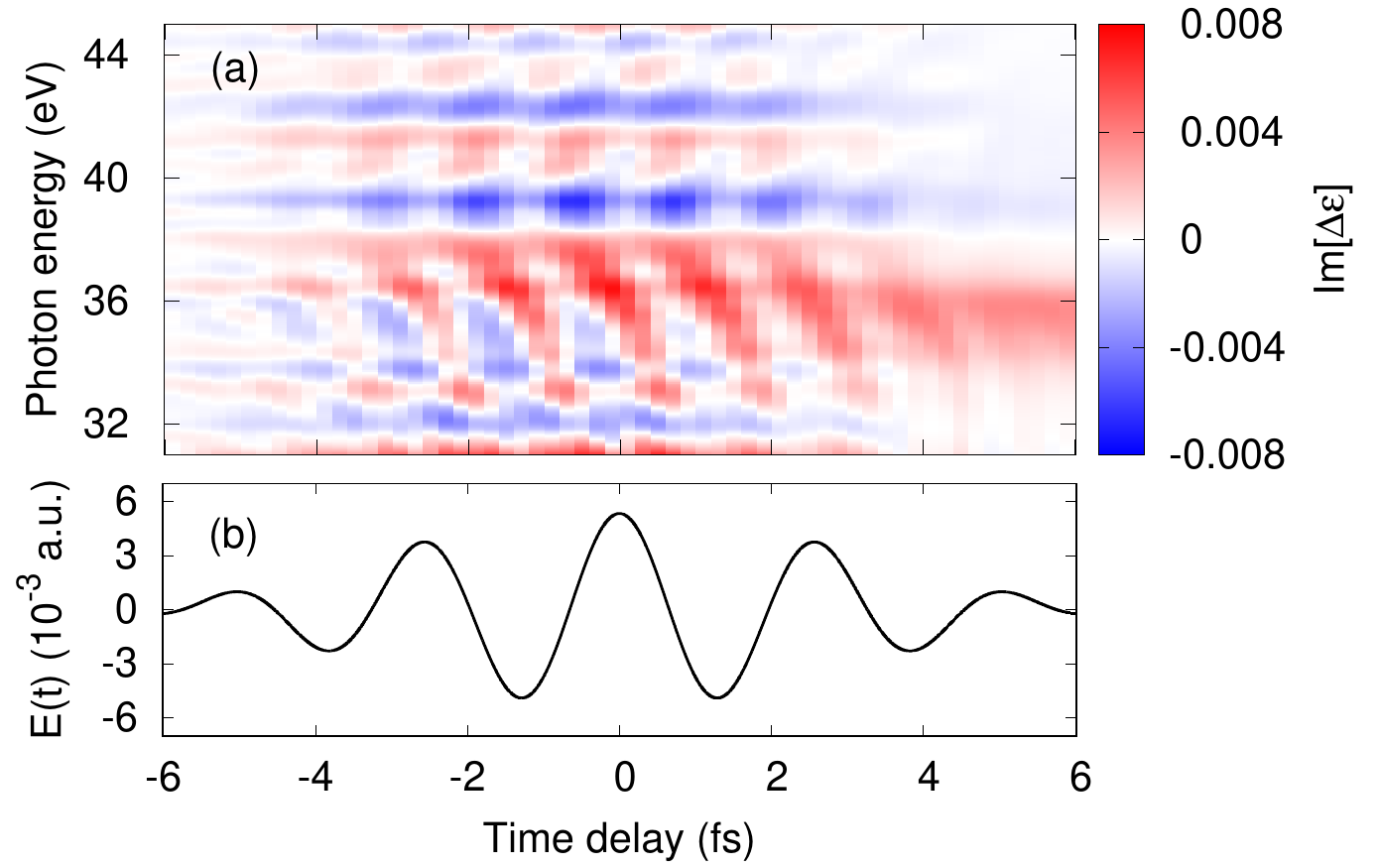}
\caption{\label{fig:tr_eps_GaAs}
(a) the imaginary part of the transient dielectric function of GaAs computed with the numerical pump-probe experiment \cite{Schlaepfer2018}. (b) the time-profile of the pump laser pulse.
}
\end{figure}

Figure~\ref{fig:tr_eps_GaAs}~(a) shows the computed transient dielectric function around the absorption edge, $38$~eV. In Fig.~\ref{fig:tr_eps_GaAs}~(b), the time-profile of the pump electric field is also shown. The transient dielectric function shows complex behaviors: a dispersive feature is observed during the laser irradiation, while the long-lasting signal is observed around $35$~eV after the irradiation. The complex transient signal reflects the complex electronic structure of the solid, and it is not so straightforward to understand the microscopic physics behind the signal.

To provide the microscopic insight into the transient optical property, we developed the orbital decomposition of the transient response \cite{Lucchini916,Schlaepfer2018}. In the independent particle approximation, the time-dependent Hamiltonian can be instantaneously defined by the time-dependent vector potential $\bm A(t)$, which is the sum of the pump and probe vector potentials, $\bm A(t)=\bm A_{pump}(t)+\bm A_{probe}(t)$. Hence, the Hamiltonian in Eq.~(\ref{eq:tdks-velocity-bloch-ham}) can be rewritten as $\hat h_{\bm k}(t)=\hat h_{\bm k+\bm A(t)}$. Furthermore, in the pump-probe simulation, the probe field is so weak that the response to the probe can be described by the linear response. Thus, one can decompose the Hamiltonian into the pump and probe parts as
\begin{align}
\hat h_{\bm k+ \bm A(t)} &= \hat h_{\bm k+\bm A_{pump}(t)} + \frac{\hat h_{\bm k+\bm A_{pump}(t)}}{\partial \bm A_{pump}}\cdot \bm A_{probe}(t) \nonumber \\
&= \hat h_{\bm k+\bm A_{pump}(t)} + \hat H_{\bm k,probe}(t),
\end{align}
where the probe Hamiltonian $\hat H_{\bm k,probe}(t)$ is defined as
\begin{align}
\hat H_{\bm k, probe}(t)=\frac{\hat h_{\bm k+ \bm A_{pump}(t)}}{\partial \bm A_{pump}}\cdot \bm A_{probe}(t).
\end{align}
Although the whole process of the numerical pump-probe experiment is the nonlinear process, it can bee seen as a linear response of the pumped system, which is described by the pump Hamiltonian $\hat h_{\bm k+\bm A_{pump}(t)}$, respect to the probe perturbation $\hat H_{\bm k, probe}(t)$. Hence, the pump-probe simulation can be seen as a linear response calculation with the weak probe Hamiltonian $\hat H_{\bm k, probe}(t)$. Based on this fact, one can decompose the response to the probe perturbation by decomposing the probe Hamiltonian itself, resulting in the decomposition of the response function.

To practically perform the orbital decomposition, we introduce instantaneous eigenstates of the pump Hamiltonian as,
\begin{align}
\hat h_{\bm k + \bm A_{pump}(t)} \left | u^H_{b\bm k}(t) \right \rangle =\epsilon^H_{b\bm k}(t)\left | u^H_{b\bm k}(t) \right \rangle,
\end{align}
where $\left | u^H_{b\bm k}(t) \right \rangle$ are the instantaneous eigenstates, and $\epsilon^H_{b\bm k}(t)$ are the corresponding eigenvalues. With these orbitals, one can construct a decomposed probe Hamiltonian as
\begin{align}
\hat H^{decomp}_{\bm k, probe}(t)= \sum_{b b'} w_{bb'} \left | u^H_{b\bm k}(t)\right \rangle \left \langle u^H_{b\bm k}(t) \right |
 \hat H_{\bm k, probe}(t)
\left | u^H_{b'\bm k}(t)\right \rangle \left \langle u^H_{b'\bm k}(t) \right |,
\label{eq:decomp-probe-ham}
\end{align}
where $w_{bb'}$ is weight of the corresponding transition. By setting $w_{bb'}$ to $1$, the decomposed Hamiltonian includes the transition between two bands, $b$ and $b'$. In contrast, by setting $w_{bb'}$ to $0$, the decomposed Hamiltonian excludes the corresponding transition. Note that, if all the weight $w_{bb'}$ is set to $1$, the decomposed probe Hamiltonian in Eq.~(\ref{eq:decomp-probe-ham}) recovers the original probe Hamiltonian $\hat H_{\bm k, probe}(t)$.

\begin{figure}[htb]
\centering\includegraphics[width=1.00\linewidth]{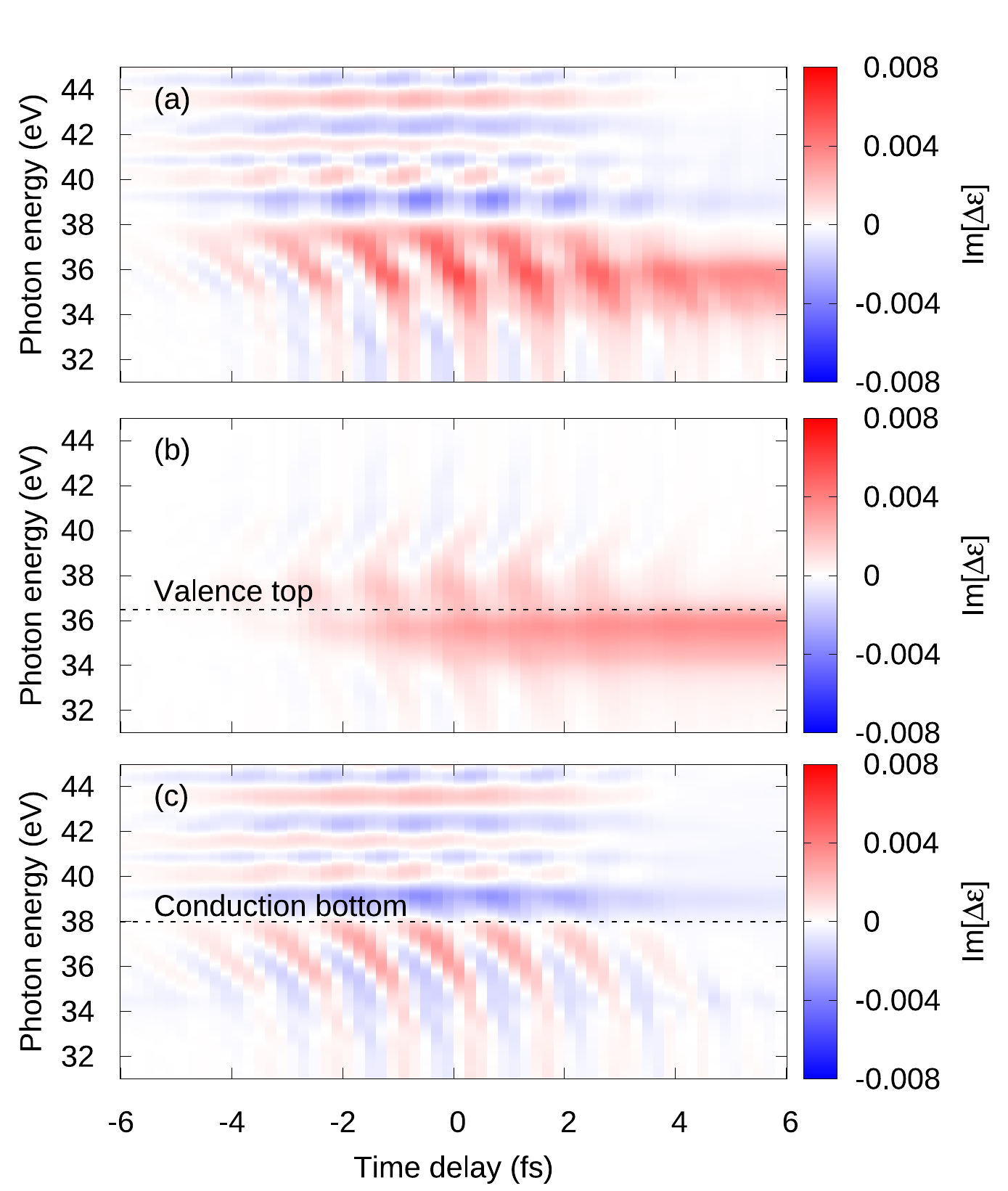}
\caption{\label{fig:tr_eps_GaAs_decomp}
Decomposed transient dielectric function of GaAs. The contributions of (a) the transition from As-$3d$ levels, (b) the transition between As-$3d$ levels and the valence bands, and (c) the transition between As-$3d$ levels and the conduction bands, are shown.
}
\end{figure}

In the pump-probe simulation, the probe field is weak enough to realize the linear response. Thus, by decomposing the Hamiltonian with Eq.~(\ref{eq:decomp-probe-ham}), the induced current can be decomposed accordingly. Furthermore, the optical conductivity, which is directly defined by the induced current in Eq.~(\ref{eq:pump-probe-sigma}), can also be decomposed by the probe Hamiltonian decomposition. Figure~\ref{fig:tr_eps_GaAs_decomp}~(a) shows the transient dielectric function computed with the decomposed probe Hamiltonian that consists of only the transition from As-$3d$ states. Comparing the decomposed result in Fig.~\ref{fig:tr_eps_GaAs_decomp}~(a) with the full result in Fig.~\ref{fig:tr_eps_GaAs}~(a), one can confirm that the transient absorption of GaAs is dominated by the transition from As-$3d$ states.

To further study the microscopic origin of the transient absorption of GaAs in Fig.~\ref{fig:tr_eps_GaAs_decomp}~(a), we decompose the transient dielectric function into more detailed components. Figure~\ref{fig:tr_eps_GaAs_decomp}~(b) shows the contribution from the transition between the As-$3d$ states and the the valence bands, and Figure~\ref{fig:tr_eps_GaAs_decomp}~(c) shows the contribution from the transition between the As-$3d$ states and the conduction bands (see the band structure in Fig.~\ref{fig:GaAs_lin}~(c)).

In Fig.~\ref{fig:tr_eps_GaAs_decomp}~(b), the transition energy between the As-$3d$ level and the valence top is shown as the horizontal black-dotted line. The transient dielectric function in Fig.~\ref{fig:tr_eps_GaAs_decomp}~(b) shows the long-lasting signal below the horizontal black-dotted line. This long-lasting signal can be straightforwardly understood by the state-filling effect: before the pump laser irradiation, the valence bands are fully occupied. Thus, the transition from the semi-core levels to the valence bands are forbidden by the Pauli blocking. Once the pump pulse is irradiated, electrons are excited from the valence bands to the conduction bands, and holes are created in the valence bands. As a result, the transition from the semi-core levels to the valence bands are allowed, and it causes the enhancement of the absorption. 

In Fig.~\ref{fig:tr_eps_GaAs_decomp}~(c), the transition energy between the As-$3d$ level and the conduction top is shown as the horizontal black-dotted line. As seen from the figure, the transient dielectric function shows the transient signal that exists only during the pump irradiation but disappears after the irradiation. Remarkably, one can observe the dispersive signal below the transition energy between the As-$3d$ levels and the conduction bottom, although the signal comes from the transition between the semi-core states and the conduction bands. In the previous work \cite{Schlaepfer2018}, we further analyzed this transient signal with a simplified three-band model on the basis of the first-principles electronic structure and clarified that the transient signal in Fig.~\ref{fig:tr_eps_GaAs_decomp}~(c) originates from the pump-induced intraband transitions, indicating the real-time observation of the dynamical Franz-Keldysh effect \cite{PhysRevLett.76.4576,Lucchini916}.

\section{Summary \label{sec:summary}}

In this article, we reviewed the recent development of the first-principles analysis of light-induced electron dynamics in solids. The real-time electron dynamics simulation is a powerful tool to analyze the highly-nonlinear ultrafast phenomena in the time domain and the linear optical properties of solids. We first reviewed the basic theoretical framework based on the time-dependent density functional theory (TDDFT), which offers a formally exact description of the dynamics of many-body systems. Then, we described the practical implementation to numerically solve the time-dependent Kohn-Sham equation, which is our central equation of motion to describe the electron dynamics. Here, the two numerical approaches were introduced: one is the real-space grid representation, and the other is the basis expansion approach. One can choose one of the methods, depending on the computational costs and the nature of dynamics. The real-space grid representation can capture complex wavefunction propagation as it consists of a large degree of freedom. However, due to the large degree of freedom, the real-space grid representation tends to require higher computational costs, and it may become infeasible for the (semi) core electron dynamics. In contrast, the basis expansion approach may describe the electron dynamics with smaller computational costs since the basis set is prepared such that the basic transitions in solids, intraband and interband transitions, are naturally described.

Then, we reviewed the practical applications of the electron dynamics simulation to the attosecond spectroscopy by demonstrating the basic workflow of the first-principles analysis. First, we revisited the application to the attosecond transient absorption spectroscopy for Titanium \cite{Volkov2019}. There, we reviewed the linear response calculation for solids in the time domain to compute the linear optical property, such as dielectric function. Then, we further reviewed the two extensions of the linear response calculation: one is the numerical pump-probe experiment, where the experimental pump-probe technique is mimicked by the first-principles simulation. The other is the linear response calculation with the finite electron temperature. While the numerical pump-probe experiment provides the transient optical property in the nonequilibrium phase, the linear response calculation with the finite temperature provides the optical property in a quasi-equilibrium phase, where the electron temperature is much higher than the lattice (phonon) temperature.

The first-principle electron dynamics simulation is useful not only for the reproduction of the experimental results but also for clarification of microscopic physics behind the experimental observation. To provide microscopic insight into the phenomena, we developed the two microscopic decompositions of the simulation results. One is the mechanism-based decomposition: here, we decompose the optical property of excited systems into the three physical mechanisms: (i) state-filling effect, (ii) the local-field effect, and (iii) the band-structure renormalization. The mechanism-based decomposition identified the origin of the light-induced modification of Titanium as the modification of the microscopic screening through the light-induced electron localization \cite{Volkov2019}. The other microscopic decomposition is the band decomposition of the transient optical property. This decomposition is performed by the decomposition of the probe perturbation in the numerical pump-probe experiment. The band decomposition has been applied to the attosecond transient absorption spectroscopy of GaAs and clarified that the light-induced modification of the optical property is dominated by the two kinds of transitions: one is the transitions between the As-$3d$ band and the valence bands, and the other is the transitions between the As-$3d$ band and the conduction bands. In the previous work, the simple model analysis has been done on the basis of the band decomposition, and it has been clarified that the light-induced intraband transitions play a significant role in the modification of the optical properties of semiconductors under the presence of resonant driving light \cite{Schlaepfer2018}.

For the moment, the experimental accessibility to nano-scale and femtosecond-scale information is very limited. In contrast, the first-principles approach enables us to access such micro and ultrafast spatiotemporal scales, while still, it is not straightforward to access the macroscopic outcome observed in the experiment. Thus, the first-principles simulations and experiments have complementary roles in studying the ultrafast phenomena, and the combination of the two approaches is key to uncover the physics behind.

\section*{Acknowledgement}
The author acknowledges fruitful discussions with A.~Rubio, K.~Yabana, U.~Keller, L.~Gallmann, M.~Lucchini, M.~Volkov, and F.~Schlaepfer. This work was supported by JSPS KAKENHI Grant Number 20K14382.


\bibliographystyle{elsarticle-num-names}
\bibliography{ref.bib}







\end{document}